\documentclass[fleqn,usenatbib]{mnras}

\usepackage[T1]{fontenc}
\usepackage{ae,aecompl}

\usepackage{graphicx}
\usepackage{amsmath}
\usepackage{amssymb}
\usepackage{listings}

\title[A visualization toolkit for the simulation code {\sc arepo}]{A visualization toolkit for the simulation code {\sc arepo}}

\author[T. H. Greif et al.]{
Thomas H. Greif,$^{1,2}$\thanks{E-mail: tgreif@uni-heidelberg.de}
\\
$^{1}$Universit\"{a}t Heidelberg, Zentrum f\"{u}r Astronomie, Institut f\"{u}r theoretische Astrophysik,\\ Albert-Ueberle-Str. 2, 69120 Heidelberg, Germany\\
$^{2}$Institute for the Physics and Mathematics of the Universe, 5-1-5 Kashiwanoha, Kashiwa, 277-8583, Japan\\
}

\pubyear{2018}

\begin{document}
\label{firstpage}
\pagerange{\pageref{firstpage}--\pageref{lastpage}}
\maketitle

\begin{abstract}
In this paper I describe the visualization toolkit {\sc sator}, which is designed to read, analyze and visualize simulation data of the moving-mesh code {\sc arepo}. It is written in Python and employs a graphical user interface based on the Tkinter module, providing an interactive and intuitive user experience. All three snapshot formats employed by {\sc arepo} are supported, including the HDF5 format. Individual snapshot fields, for example cell coordinates, are read interactively only when they are needed. Refined fields used for plots, for example the temperature, are constructed from snapshot fields and can be introduced with only a few lines of code. {\sc sator} currently supports the generation of image slices and projections, which can be zoomed, moved and rotated with minimal computational effort. Furthermore, various phase space and line plots can be created. Due to its modular nature, additional analysis and plotting routines can be easily implemented and are under construction. The code is well documented and is intended to be free and open source, such that any member of the astrophysical community may contribute.
\end{abstract}


\section{Introduction}

The complexity of multi-dimensional simulations and the number of resolution elements employed has steadily increased in the last few decades \citep{teyssier02, springel05a, springel10a, hayes11, mignone12, bryan14, hopkins15}. This has driven a need for increasingly sophisticated tools to read, analyze and visualize the simulation data. This data is typically stored in so-called snapshots, which are written in periodic intervals and consist of a set of physical fields, for example the density and temperature, of all resolution elements. Since the size of a snapshot scales with the number of resolution elements employed, the tools to read and analyze the simulation data must evolve hand in hand with the simulation codes to maintain an efficient post-processing pipeline. An example of such a development is the analysis toolkit {\sc yt} for the simulation code {\sc enzo} \citep{turk11b}. For the moving-mesh code {\sc arepo}, which has been used to investigate a number of astrophysical problems \citep{greif11a, pbs11, bs12, vogelsberger12, munoz13, pakmor13, mps14, mocz14, vogelsberger14, becerra15, schaal15, kannan16, ohlmann16, grand17, pfrommer17, weinberger17, marinacci18, mckinnon18, pillepich18}, such a framework is still missing.

I here present the toolkit {\sc sator}, which is designed to read, analyze and visualize output of the simulation code {\sc arepo}. It is written mainly in Python, with some computationally intensive routines wrapped in C functions. {\sc sator} employs a graphical user interface (GUI) based on the Tkinter module, which increases its usability compared to tools that rely solely on scripts to create individual plots. Nevertheless, {\sc sator} can also be run in script mode with minimal code overhead to create customized plots. The code is highly modular, allowing for easy extensions of the user interface to create more plot options. All three currently supported snapshots formats of {\sc arepo} are supported, including the HDF5 format.

In the following sections, I describe the steps necessary to install the toolkit from the Bitbucket repository, the functionality of the GUI, the reading methods for snapshots, and the different types of fields employed. I then discuss the various plots that can be generated, how simple scripts can be used to created customized plots, and possible improvements to the code that can be implemented in the near future.

\begin{figure*}
\begin{center}
\includegraphics[width=16cm]{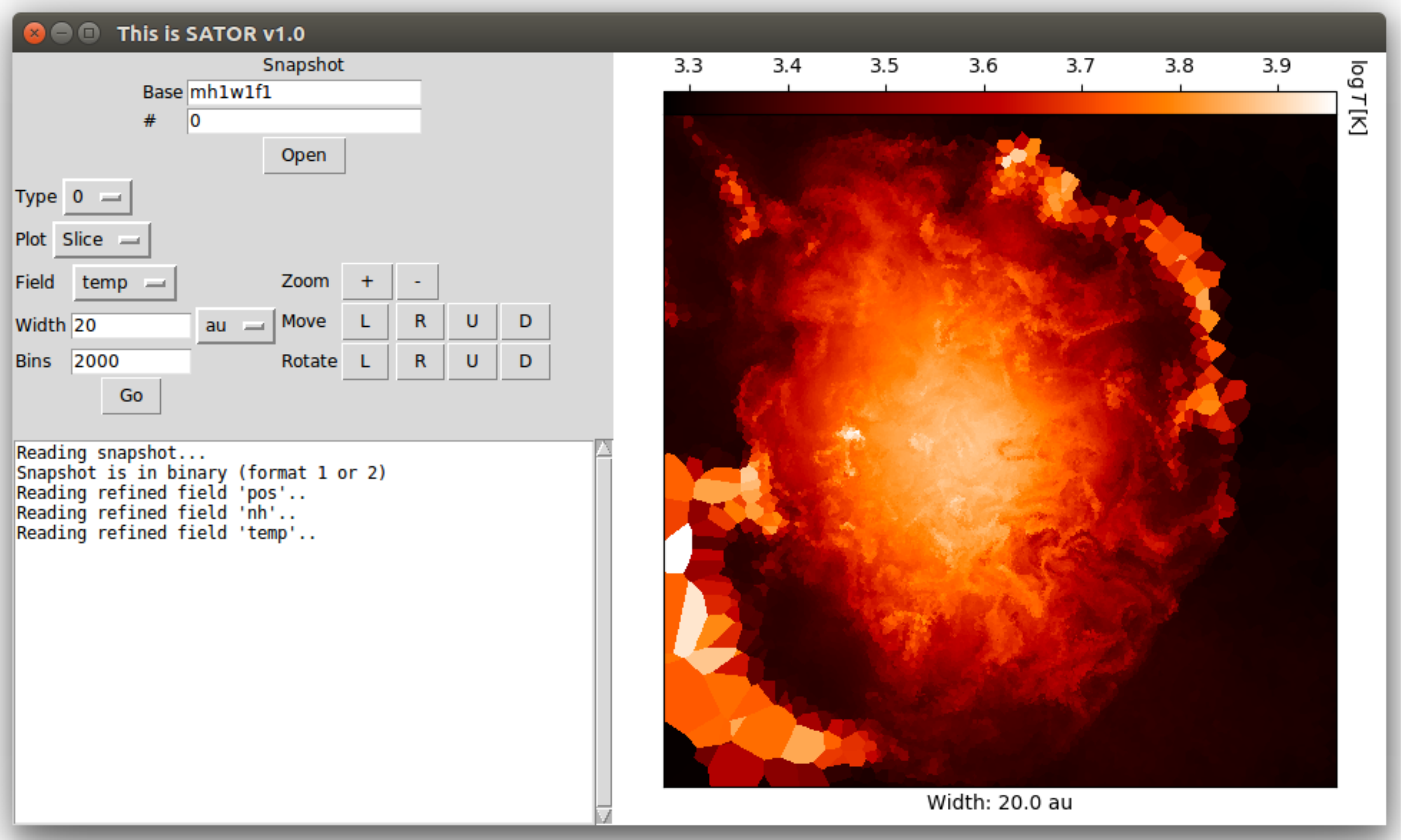}
\caption{The user interface of {\sc sator} for an image slice through the center of the simulation box. The menu frame consists of fields for the name and number of the snapshot, the selected type of resolution element, the target field (in this case the temperature), the width of the image, and the number of pixels. Furthermore, buttons to quickly zoom, move and rotate the image are available. The bottom left frame shows various informational messages as the snapshot data is read.}
\label{slice}
\end{center}
\end{figure*}
\section{Methodology}

\subsection{Preparation}
Most of the code is written in Python and uses a number of additional Python modules: NumPy, SciPy, Matplotlib, Tkinter, H5py, Ctypes, Struct, Sys, Os and ScrolledText. These are typically included in a default installation of Python, so that no additional steps are necessary. The source code of {\sc sator} is stored in a Bitbucket repository, which can be cloned from \texttt{git@bitbucket.org:tgreif/sator.git} once a Bitbucket account has been created and access to the repository is granted. The repository consists of a template for the parameter file, which contains customizable variables that are read during the initialization of {\sc sator}, the Python program files, a Makefile, the C libraries, and a bash script. {\sc sator} is initialized by executing the bash script with the parameter file as the first argument. The \texttt{DATADIR} environment variable must also be set to point to the directory that contains the simulation output. Various customizable settings of {\sc sator} are stored in the file \texttt{settings.py}. Initially, these do not need to be changed.

\subsection{User interface}

The entire functionality of {\sc sator} is encapsulated in a single class, the \texttt{sator} class, an instance of which is called during the start-up of {\sc sator}. The GUI is created with the Python module Tkinter, which is a thin, object-oriented layer on top of Tcl/Tk. The {\sc sator} class draws a root frame with the method \texttt{tk.Tk()} during its initialization, inside of which all other frames are placed. The root frame is then divided into three child frames: the menu frame, which contains the controls of the GUI, the output frame, which displays various information and error messages, and the plot frame, which is reserved for visualization of the data. The menu frame contains a number of child frames, the first of which is the snapshot frame, which displays options for reading the header of a snapshot. Below is an example of the code used to create a frame:
\begin{verbatim}
self.snap_frame = tk.Frame(self.menu_frame)
self.snap_frame.grid(row = 0, column = 0)
\end{verbatim}
In this case, the child frame \texttt{self.snap\_frame} is created from the parent frame \texttt{self.menu\_frame}, and the \texttt{grid} method is called to place the frame in the top left corner.

\subsection{Snapshots}
The menu frame initially displays two options to read the header of a snapshot: the name of the snapshot and its number. The header contains essential information about the snapshot, for example the physical units of the data, and the number of resolution elements. {\sc sator} automatically detects the format of the snapshot, whether it is located in a subdirectory, and among how many files the snapshot is distributed. It should be noted that {\sc arepo} supports three snapshot formats, mainly for historical reasons. In format 1, the snapshot data is stored in a binary file. A major disadvantage of this format is that one must know in advance which fields are present, how they are ordered, and for which types of resolution elements (dark matter/gas/stars) they are present. In addition, key parameters of the simulation and the configuration options are not stored with the snapshot, and must therefore be read from separate files. In format 2, the snapshot is also stored in a binary file, but each field is preceded by a variable that specifies the name of the field. This alleviates some of the problems of format 1, but still has the problem that key parameters of the simulation are not included. Format 3 alleviates all of the aforementioned problems by storing the snapshot in the HDF5 format. The fields are stored as HDF5 datasets in groups named for example \texttt{PartType0} for particle type $0$, and the auxiliary data of the simulation is stored in the groups \texttt{Header}, \texttt{Parameters} and \texttt{Config}.

When a snapshot is opened, all relevant auxiliary information of a simulation is stored in an instance of the class \texttt{auxiliary}. For snapshot format 1, the order of the fields that exist in the snapshot are read from a file specified in the variable \texttt{SnapFieldsFile} in the parameter file. A template for this file is included in the repository, and consists of three elements for each snapshot field that is present. The first column specifies the name of the field, which must be identical to the name it would have in an HDF5 file, the second column gives the number of entries for each resolution element, and the last column a string that represents the NumPy data type. The order of the fields is also important, since the fields are stored sequentially in the snapshot. For snapshot formats 2 and 3, the contents of this file are not used, since the fields and their properties are contained in the snapshot and may be directly copied to the \texttt{auxiliary} instance.

\begin{figure*}
\begin{center}
\includegraphics[width=16cm]{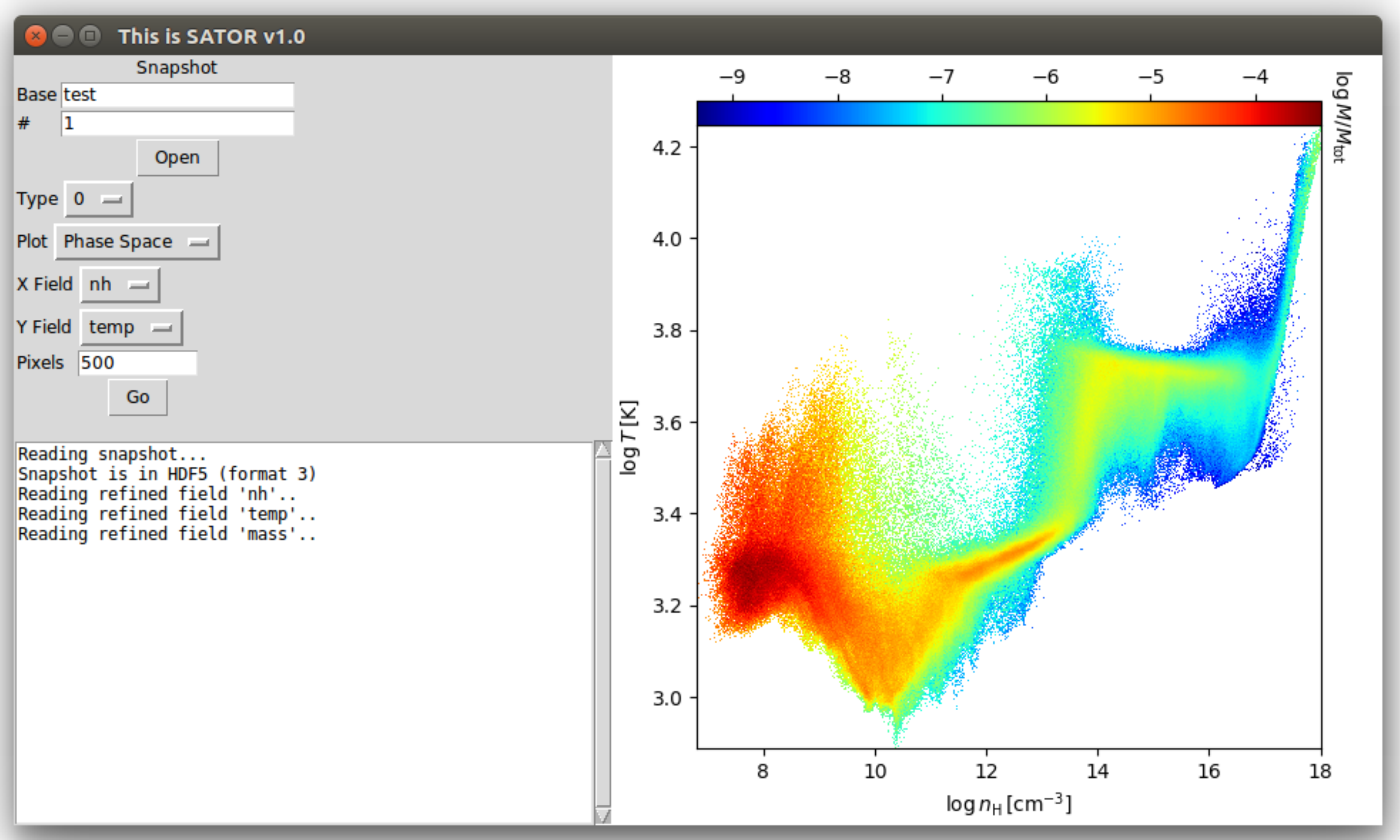}
\caption{The user interface of {\sc sator} for a phase space plot. Next to the snapshot name, number and selected type of resolution element, the menu frame shows the target fields in the x- and y-directions, next to the number of pixels. Here, the plot that is created on the right shows the temperature versus density of all resolution elements in a simulation of a minihalo, with the fractional mass in each bin color-coded according to the panel at the top.}
\label{pspace}
\end{center}
\end{figure*}

\subsection{Types and fields}

Following the storage of the auxiliary data, the types frame is created as a child frame of the menu frame. This frame contains a menu for the selection of the type of resolution element to be read. Once a particle type has been selected, variables that store information concerning the snapshot fields are initialized. A Python dictionary is used to allow quick look-up of the dataset associated with the snapshot field based on its HDF5 name (for example \texttt{Coordinates}). In addition, the number of entries per resolution element and the NumPy data type of the snapshot field are stored.

In a second step, so-called refined fields are intialized. These fields are typically computed from one or more snapshot fields, and are typically used in plots. An example for a refined field is the temperature, which is computed from the snapshot fields \texttt{Gamma}, \texttt{ChemicalAbundances} and \texttt{InternalEnergy}. The refined fields that can in principle be used are customizable via a Python list in the settings file. For example, the temperature can be added with the following list element:
\begin{verbatim}
['temp', 1, 'float64']
\end{verbatim}
Here, the first column is a user-defined name for the field, the second column specifies the number of entries per resolution element, and the third field determines the NumPy data type. The calculation of a refined field from the snapshot fields is specified in the method \texttt{get\_refined\_field()} in \texttt{fields.py}. Using again the example of the temperature, the following code section shows the calculation of the temperature field from the corresponding snapshot fields:
\begin{verbatim}
snap_field = 'Gamma'

flag, gamma = self.get_snap_field(snap_field)

if flag:
    gamma = gamma_adb

snap_field = 'ChemicalAbundances'

flag, chem = self.get_snap_field(snap_field)

if flag:
    mu = mu_prim
else:
    abhm = chem[:, 0]
    abh2 = chem[:, 1]
    abhii = chem[:, 2]
    abe = abhii
    mu = (1. + 4. * abhe)
        / (1. + abhe - abh2 + abe)

snap_field = 'InternalEnergy'

flag, u = self.get_snap_field(snap_field)

if flag:
    self.snap_field_error(snap_field)
else:
    fac = unit_energy / unit_mass
    u *= fac
    temp = mu * (gamma - 1.) * mh / kb * u
\end{verbatim}
In this example, the snapshot field \texttt{Gamma} is first read. If this field exists, it is read from the snapshot file and stored in the NumPy array \texttt{gamma}. If it does not exist, \texttt{gamma} defaults to the value for a monatomic gas. Next, the field \texttt{ChemicalAbundances} is read. If it exists, the mean molecular weight for a variable chemical composition is computed, otherwise the default value for a pure hydrogen / helium gas is used. Finally, the field \texttt{InternalEnergy} is read and together with \texttt{gamma} and \texttt{mu} used to compute the temperature of the resolution elements.

Similar to the snapshot fields, a Python dictionary for quick look-up of the dataset associated with the name of the refined field is stored for each refined field specified in the settings file, next to the number of entries per resolution element and the NumPy data type. Both types of fields are only read if they are not already present, which is indicated by a non-empty dictionary element. This substantially reduces the computational overhead. Finally, we note that while the snapshot fields are stored in code units, the refined fields are stored in cgs units.

\subsection{Plot options}

Once the snapshot and refined fields have been initialized, a child frame is created that contains a menu of the available plot options. These may be specified in the Python list \texttt{sator.plot\_options} in the settings file. If a choice for a plot option has been made, an option-specific child frame for plot sub-options is created that contains all sub-options of the specified plot option. This list may customized by modifying the method \texttt{sator.create\_plot\_sub\_options\_frame()} in \texttt{sator.py}. As an example, the sub-options of the image slice include the width of the sub-box for which the image is created and the number of pixels in the x and y directions.

\subsection{Images}
One of the main functionalities of \texttt{sator} is the possibility to create images of the fields. Currently, this is only possible for resolution elements of type 0 (gas), but a more versatile routine is under construction. The two plot options available for images are slices and projections. A slice computes the values of a field in a plane of the simulation box, while a projection shows the average value of a field along the line of sight, weighted by a user-defined function. Both plot options include sub-options to specify the field to be shown (for example the temperature), the width of the sub-box, and the number of pixels used in the x and y directions. The length units available for the width of the sub-box can be specified in the settings file.

The first time an image for a selected type of resolution element and plot option is created, a refined field is initialized that reads in the positions of all resolution elements in the snapshot. For slices, the Python method \texttt{spatial.cKDTree(pos)} is invoked, where \texttt{pos} is the refined field containing the positions of the resolution elements. This method constructs a tree for rapid look-up of the nearest neighbors of a list of evenly spaced points in a plane. This exploits the convenient property of Voronoi tesselations that a given point in space lies in the Voronoi cell of the mesh-generating point it is closest to. For projections, the subset of resolution elements that lie within the selected sub-box are extracted. These resolution elements are mapped onto a rectangular grid, where the indices of the pixels they contribute to are found by using the approximate size of a Voronoi cell, $h=(3V/(4\pi))^{1/3}$, where $V$ is the volume. This loop is the most computationally intensive step and is therefore performed in a C subroutine:
\begin{verbatim}
void projection(double* x, double* y,
double* rho, double* h, double* field,
int npart, double* proj, double* sums,
int npixels)
{
  int n, i, j, idx;
  int min1, max1, min2, max2;
  double psum, pproj;

  for(n = 0; n < npart; n++)
    {
      min1 = (x[n] + 0.5 - h[n]) * npixels;
      max1 = (x[n] + 0.5 + h[n]) * npixels;

      min2 = (y[n] + 0.5 - h[n]) * npixels;
      max2 = (y[n] + 0.5 + h[n]) * npixels;

      psum = rho[n] * rho[n];
      pproj = psum * field[n];

      for(i = min1; i <= max1; i++)
        for(j = min2; j <= max2; j++)
          {
            idx = i * nbins + j;

            proj[idx] += pproj;
            sums[idx] += psum;
          }
    }
}
\end{verbatim}
Here, \texttt{x} and \texttt{y} denote the coordinates of the resolution elements, \texttt{h} their radius, \texttt{rho} their density, and \texttt{field} the field that is to be plotted. In this case the target field is weighted with the density squared along the line of sight, which emphasizes dense regions.

For images, buttons to zoom, move and rotate the sub-box are also available. When the zoom buttons are invoked, a sub-box with a correspondingly increased or decreased width is selected, and the above steps are repeated, with the exception that for slices the tree does not need to be reconstructed. Since the existing tree contains all resolution elements of the simulation box, the tree is simply walked for a new set of grid points. This greatly speeds up the calculation.

Similarly, the move button centers the sub-box on a different point and repeats the above steps. Finally, the sub-box may be rotated along the azimuthal and polar angles around its center. For all three operations, the necessary calculations are performed on existing fields, so that no additional data must be read from the snapshot. The zoom factor, move fraction, and incremental rotation angle may be specified in the settings file.

\begin{figure}
\begin{center}
\includegraphics[width=8cm]{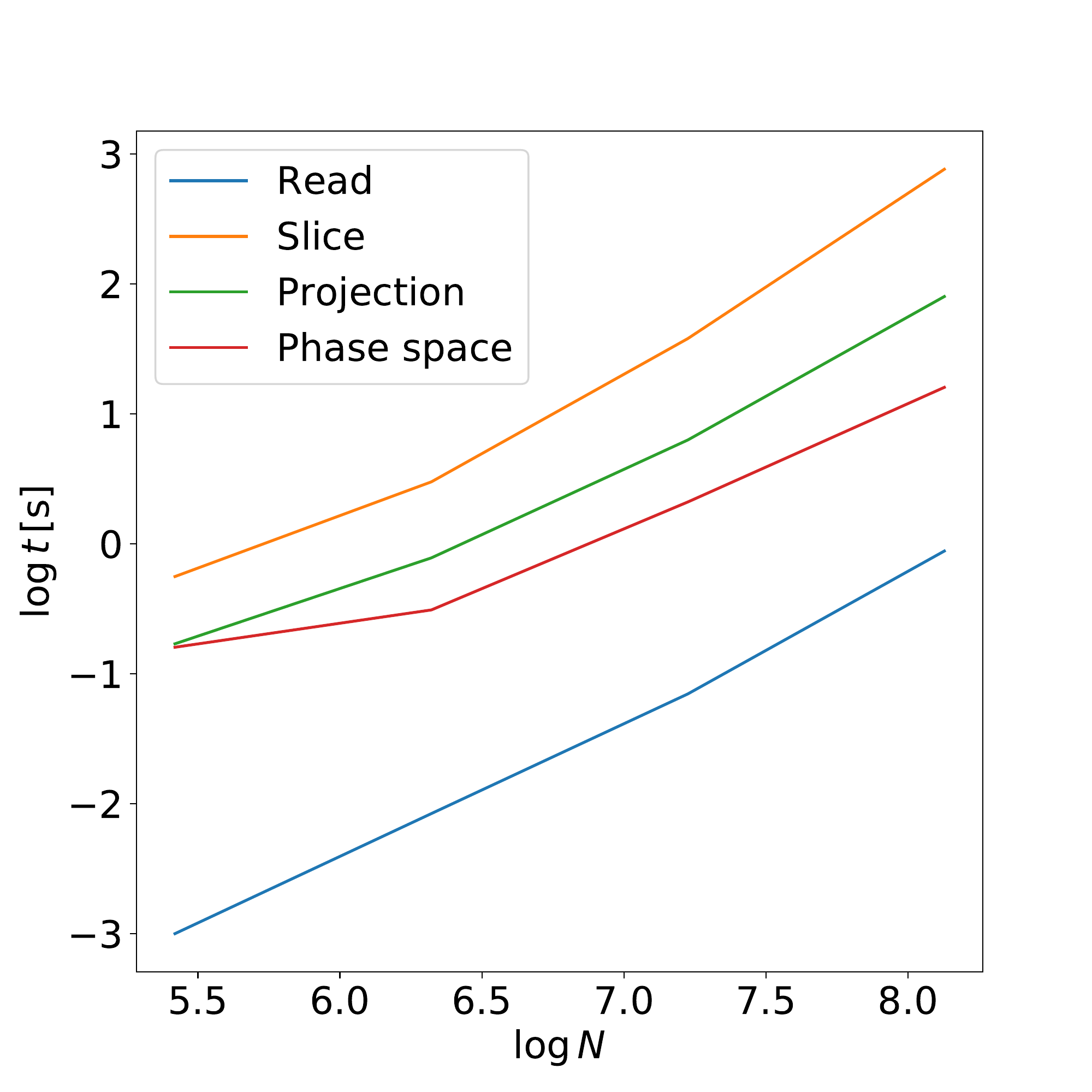}
\caption{a}
\label{performance}
\end{center}
\end{figure}

\subsection{Phase Space and Lines}

\subsection{Scripts}
{\sc sator} may also be run in a script mode, which allows the customization of plots. Below is an example of a simple script that generates a customized phase space plot:
\begin{verbatim}
from sator import *

sator = sator('par.txt')

sator.get_header('test', '1')

sator.init_fields(0)

sator.init_figure()

axes = sator.get_pspace('nh', 'temp', 500)

axes.do_something()

sator.save_figure('out.pdf')
\end{verbatim}
In this case, a sator instance is created using \texttt{par.txt} as the parameter file. Next, the header of a snapshot named \texttt{test} and snapshot number 1 is read, and the fields are initialized for particle type 0. After the figure has been initialized, the phase space routine is called, using the density and temperature as x and y axis, respectively, next to the number of pixels. This routine returns an \texttt{Axe}s instance of the Matplotlib module, which can then be used to customize the plot. Finally, the figure is saved to the file \texttt{out.pdf}.

\subsection{Performance}
In Figure~3, we show the performance of {\sc sator} for various tasks. The tests were performed on an Intel Core i7-7500 CPU with 16 GB memory. The blue line shows the execution time for a field with one entry versus the number of resolution elements, the orange line the execution time for an image slice of the entire simulation box, the green line for a projection, and the red line for a phase space plot. Evidently, the performance of {\sc sator} is not limited by the reading of snapshot fields, which merely takes about $1\,{\rm s}$ for about $10^8$ resolution elements. Instead, the performance is mainly limited by the routines to generate plots. The main bottleneck for image slices is the construction of the tree, and for projections and phase space plots the binning of the individual particles into the image pixels, which is implemented with calls to C functions. For $10^7$ resolution elements, the generation of an image slice takes about $10\,{\rm s}$, a projection about $3\,{\rm s}$, and a phase space plot about $1\,{\rm s}$. However, the performance decreases by an order of magnitude once $10^8$ resolution elements are approached. In the current serial implementation of {\sc sator}, it is therefore not feasible to run {\sc sator} on snapshots with $10^8$ or more resolution elements.

\subsection{Future directions}

A number of possibilities exist to further improve {\sc sator}. First, the image generation routine should be augmented by the capability to plot other types of resolution elements as well, for example dark matter or stars. A routine for volume rendering of the simulation data by casting rays through the domain would allow for a better visualization of the three-dimensional morphology of structures. The output of the friends-of-friends halo finder and of Subfind, which are part of {\sc arepo}, are stored in separate files and could in principle also be read with {\sc sator}. This would be advantageous for cosmological simulations, where the halo data can be used to only read resolution elements that are part of a specific halo. Another improvement that is likely to be implemented in the near future is the parallelization of the code with the Python MPI library mpi4py. This would be useful in particular for snapshots with more than $10^8$ resolution elements, where the runtime currently approaches $100\,{\rm s}$. Another constraint is the limited amount of memory available on a single node. Both of these problems can be addressed by an efficient parallelization strategy. 

\section{Summary and conclusions}
I have introduced the Python toolkit {\sc sator}, which is designed to read, analyze and visualize simulation data output by the simulation code {\sc arepo}. All three snapshot formats of {\sc arepo} are supported, including the easily readable HDF5 format. It employs a graphical user interface for an intuitive user experience, but can also be used in a script-based mode to create customized plots. In the current framework, {\sc sator} supports the generation of image slices and projections, as well as various phase space and line plots. Data is only read when necessary and when it is not already contained in memory. For example, the generation of different image slices of the same snapshot is computationally expensive only during the initial execution. Subsequent views rely on existing data and are therefore much cheaper.

{\sc sator} is highly modular and allows a straightforward inclusion of additional fields, which are typically constructed from raw snapshot fields. Additional plot options can be added to the user interface based on existing templates. The performance of {\sc sator} is good for up to $10^6$--$10^7$ resolution elements, but becomes unsatisfactory for about $10^8$ resolution elements. Work is currently underway to fully parallelize the underlying routines for reading and binning data. The toolkit is intended to be free and open source, and is available online via a Bitbucket repository. Permission to access the repository may be granted by sending an email to the author.

\section*{Acknowledgements}
THG would like to acknowledge support from the Heidelberg Institute for Theoretical Studies, where part of this work was completed. THG also thanks the European Research Council for funding in the ERC Advanced Grant STARLIGHT (project number 339177).

\bibliographystyle{mnras}

\begin{thebibliography}{}
\makeatletter
\relax
\def\mn@urlcharsother{\let\do\@makeother \do\$\do\&\do\#\do\^\do\_\do\%\do\~}
\def\mn@doi{\begingroup\mn@urlcharsother \@ifnextchar [ {\mn@doi@}
  {\mn@doi@[]}}
\def\mn@doi@[#1]#2{\def\@tempa{#1}\ifx\@tempa\@empty \href
  {http://dx.doi.org/#2} {doi:#2}\else \href {http://dx.doi.org/#2} {#1}\fi
  \endgroup}
\def\mn@eprint#1#2{\mn@eprint@#1:#2::\@nil}
\def\mn@eprint@arXiv#1{\href {http://arxiv.org/abs/#1} {{\tt arXiv:#1}}}
\def\mn@eprint@dblp#1{\href {http://dblp.uni-trier.de/rec/bibtex/#1.xml}
  {dblp:#1}}
\def\mn@eprint@#1:#2:#3:#4\@nil{\def\@tempa {#1}\def\@tempb {#2}\def\@tempc
  {#3}\ifx \@tempc \@empty \let \@tempc \@tempb \let \@tempb \@tempa \fi \ifx
  \@tempb \@empty \def\@tempb {arXiv}\fi \@ifundefined
  {mn@eprint@\@tempb}{\@tempb:\@tempc}{\expandafter \expandafter \csname
  mn@eprint@\@tempb\endcsname \expandafter{\@tempc}}}

\bibitem[\protect\citeauthoryear{{Bauer} \& {Springel}}{{Bauer} \&
  {Springel}}{2012}]{bs12}
{Bauer} A.,  {Springel} V.,  2012, \mn@doi [\mnras]
  {10.1111/j.1365-2966.2012.21058.x}, \href
  {http://adsabs.harvard.edu/abs/2012MNRAS.423.2558B} {423, 2558}

\bibitem[\protect\citeauthoryear{{Becerra}, {Greif}, {Springel}  \&
  {Hernquist}}{{Becerra} et~al.}{2015}]{becerra15}
{Becerra} F.,  {Greif} T.~H.,  {Springel} V.,   {Hernquist} L.~E.,  2015,
  \mn@doi [\mnras] {10.1093/mnras/stu2284}, \href
  {http://adsabs.harvard.edu/abs/2015MNRAS.446.2380B} {446, 2380}

\bibitem[\protect\citeauthoryear{{Bryan} et~al.,}{{Bryan}
  et~al.}{2014}]{bryan14}
{Bryan} G.~L.,  et~al., 2014, \mn@doi [\apjs] {10.1088/0067-0049/211/2/19},
  \href {http://adsabs.harvard.edu/abs/2014ApJS..211...19B} {211, 19}

\bibitem[\protect\citeauthoryear{{Grand} et~al.,}{{Grand}
  et~al.}{2017}]{grand17}
{Grand} R.~J.~J.,  et~al., 2017, \mn@doi [\mnras] {10.1093/mnras/stx071}, \href
  {http://adsabs.harvard.edu/abs/2017MNRAS.467..179G} {467, 179}

\bibitem[\protect\citeauthoryear{{Greif}, {Springel}, {White}, {Glover},
  {Clark}, {Smith}, {Klessen}  \& {Bromm}}{{Greif} et~al.}{2011}]{greif11a}
{Greif} T.~H.,  {Springel} V.,  {White} S.~D.~M.,  {Glover} S.~C.~O.,  {Clark}
  P.~C.,  {Smith} R.~J.,  {Klessen} R.~S.,   {Bromm} V.,  2011, \mn@doi [\apj]
  {10.1088/0004-637X/737/2/75}, \href
  {http://adsabs.harvard.edu/abs/2011ApJ...737...75G} {737, 75}

\bibitem[\protect\citeauthoryear{{Hayes}, {Norman}, {Fiedler}, {Bordner}, {Li},
  {Clark}, {Ud-Doula}  \& {Mac Low}}{{Hayes} et~al.}{2011}]{hayes11}
{Hayes} J.~C.,  {Norman} M.~L.,  {Fiedler} R.~A.,  {Bordner} J.~O.,  {Li}
  P.~S.,  {Clark} S.~E.,  {Ud-Doula} A.,   {Mac Low} M.-M.,  2011, {ZEUS-MP/2:
  Computational Fluid Dynamics Code}, Astrophysics Source Code Library
  (\mn@eprint {ascl} {1102.028})

\bibitem[\protect\citeauthoryear{{Hopkins}}{{Hopkins}}{2015}]{hopkins15}
{Hopkins} P.~F.,  2015, \mn@doi [\mnras] {10.1093/mnras/stv195}, \href
  {http://adsabs.harvard.edu/abs/2015MNRAS.450...53H} {450, 53}

\bibitem[\protect\citeauthoryear{{Kannan}, {Springel}, {Pakmor}, {Marinacci}
  \& {Vogelsberger}}{{Kannan} et~al.}{2016}]{kannan16}
{Kannan} R.,  {Springel} V.,  {Pakmor} R.,  {Marinacci} F.,   {Vogelsberger}
  M.,  2016, \mn@doi [\mnras] {10.1093/mnras/stw294}, \href
  {http://adsabs.harvard.edu/abs/2016MNRAS.458..410K} {458, 410}

\bibitem[\protect\citeauthoryear{{Marinacci}, {Pakmor}  \&
  {Springel}}{{Marinacci} et~al.}{2014}]{mps14}
{Marinacci} F.,  {Pakmor} R.,   {Springel} V.,  2014, \mn@doi [\mnras]
  {10.1093/mnras/stt2003}, \href
  {http://adsabs.harvard.edu/abs/2014MNRAS.437.1750M} {437, 1750}

\bibitem[\protect\citeauthoryear{{Marinacci}, {Vogelsberger}, {Kannan}, {Mocz},
  {Pakmor}  \& {Springel}}{{Marinacci} et~al.}{2018}]{marinacci18}
{Marinacci} F.,  {Vogelsberger} M.,  {Kannan} R.,  {Mocz} P.,  {Pakmor} R.,
  {Springel} V.,  2018, \mn@doi [\mnras] {10.1093/mnras/sty397}, \href
  {http://adsabs.harvard.edu/abs/2018MNRAS.476.2476M} {476, 2476}

\bibitem[\protect\citeauthoryear{{McKinnon}, {Vogelsberger}, {Torrey},
  {Marinacci}  \& {Kannan}}{{McKinnon} et~al.}{2018}]{mckinnon18}
{McKinnon} R.,  {Vogelsberger} M.,  {Torrey} P.,  {Marinacci} F.,   {Kannan}
  R.,  2018, \mn@doi [\mnras] {10.1093/mnras/sty1248}, \href
  {http://adsabs.harvard.edu/abs/2018MNRAS.478.2851M} {478, 2851}

\bibitem[\protect\citeauthoryear{{Mignone}, {Zanni}, {Tzeferacos}, {van
  Straalen}, {Colella}  \& {Bodo}}{{Mignone} et~al.}{2012}]{mignone12}
{Mignone} A.,  {Zanni} C.,  {Tzeferacos} P.,  {van Straalen} B.,  {Colella} P.,
    {Bodo} G.,  2012, \mn@doi [\apjs] {10.1088/0067-0049/198/1/7}, \href
  {http://adsabs.harvard.edu/abs/2012ApJS..198....7M} {198, 7}

\bibitem[\protect\citeauthoryear{{Mocz}, {Vogelsberger}, {Sijacki}, {Pakmor}
  \& {Hernquist}}{{Mocz} et~al.}{2014}]{mocz14}
{Mocz} P.,  {Vogelsberger} M.,  {Sijacki} D.,  {Pakmor} R.,   {Hernquist} L.,
  2014, \mn@doi [\mnras] {10.1093/mnras/stt1890}, \href
  {http://adsabs.harvard.edu/abs/2014MNRAS.437..397M} {437, 397}

\bibitem[\protect\citeauthoryear{{Mu{\~n}oz}, {Springel}, {Marcus},
  {Vogelsberger}  \& {Hernquist}}{{Mu{\~n}oz} et~al.}{2013}]{munoz13}
{Mu{\~n}oz} D.~J.,  {Springel} V.,  {Marcus} R.,  {Vogelsberger} M.,
  {Hernquist} L.,  2013, \mn@doi [\mnras] {10.1093/mnras/sts015}, \href
  {http://adsabs.harvard.edu/abs/2013MNRAS.428..254M} {428, 254}

\bibitem[\protect\citeauthoryear{{Ohlmann}, {R{\"o}pke}, {Pakmor}  \&
  {Springel}}{{Ohlmann} et~al.}{2016}]{ohlmann16}
{Ohlmann} S.~T.,  {R{\"o}pke} F.~K.,  {Pakmor} R.,   {Springel} V.,  2016,
  \mn@doi [\apjl] {10.3847/2041-8205/816/1/L9}, \href
  {http://adsabs.harvard.edu/abs/2016ApJ...816L...9O} {816, L9}

\bibitem[\protect\citeauthoryear{{Pakmor}, {Bauer}  \& {Springel}}{{Pakmor}
  et~al.}{2011}]{pbs11}
{Pakmor} R.,  {Bauer} A.,   {Springel} V.,  2011, \mn@doi [\mnras]
  {10.1111/j.1365-2966.2011.19591.x}, \href
  {http://adsabs.harvard.edu/abs/2011MNRAS.418.1392P} {418, 1392}

\bibitem[\protect\citeauthoryear{{Pakmor}, {Kromer}, {Taubenberger}  \&
  {Springel}}{{Pakmor} et~al.}{2013}]{pakmor13}
{Pakmor} R.,  {Kromer} M.,  {Taubenberger} S.,   {Springel} V.,  2013, \mn@doi
  [\apjl] {10.1088/2041-8205/770/1/L8}, \href
  {http://adsabs.harvard.edu/abs/2013ApJ...770L...8P} {770, L8}

\bibitem[\protect\citeauthoryear{{Pfrommer}, {Pakmor}, {Schaal}, {Simpson}  \&
  {Springel}}{{Pfrommer} et~al.}{2017}]{pfrommer17}
{Pfrommer} C.,  {Pakmor} R.,  {Schaal} K.,  {Simpson} C.~M.,   {Springel} V.,
  2017, \mn@doi [\mnras] {10.1093/mnras/stw2941}, \href
  {http://adsabs.harvard.edu/abs/2017MNRAS.465.4500P} {465, 4500}

\bibitem[\protect\citeauthoryear{{Pillepich} et~al.,}{{Pillepich}
  et~al.}{2018}]{pillepich18}
{Pillepich} A.,  et~al., 2018, \mn@doi [\mnras] {10.1093/mnras/stx2656}, \href
  {http://adsabs.harvard.edu/abs/2018MNRAS.473.4077P} {473, 4077}

\bibitem[\protect\citeauthoryear{{Schaal}, {Bauer}, {Chandrashekar}, {Pakmor},
  {Klingenberg}  \& {Springel}}{{Schaal} et~al.}{2015}]{schaal15}
{Schaal} K.,  {Bauer} A.,  {Chandrashekar} P.,  {Pakmor} R.,  {Klingenberg} C.,
    {Springel} V.,  2015, \mn@doi [\mnras] {10.1093/mnras/stv1859}, \href
  {http://adsabs.harvard.edu/abs/2015MNRAS.453.4278S} {453, 4278}

\bibitem[\protect\citeauthoryear{{Springel}}{{Springel}}{2005}]{springel05a}
{Springel} V.,  2005, \mn@doi [\mnras] {10.1111/j.1365-2966.2005.09655.x},
  \href {http://adsabs.harvard.edu/abs/2005MNRAS.364.1105S} {364, 1105}

\bibitem[\protect\citeauthoryear{{Springel}}{{Springel}}{2010}]{springel10a}
{Springel} V.,  2010, \mn@doi [\mnras] {10.1111/j.1365-2966.2009.15715.x},
  \href {http://adsabs.harvard.edu/abs/2010MNRAS.401..791S} {401, 791}

\bibitem[\protect\citeauthoryear{{Teyssier}}{{Teyssier}}{2002}]{teyssier02}
{Teyssier} R.,  2002, \mn@doi [\aap] {10.1051/0004-6361:20011817}, \href
  {http://adsabs.harvard.edu/abs/2002A%26A...385..337T} {385, 337}

\bibitem[\protect\citeauthoryear{{Turk}, {Smith}, {Oishi}, {Skory}, {Skillman},
  {Abel}  \& {Norman}}{{Turk} et~al.}{2011}]{turk11b}
{Turk} M.~J.,  {Smith} B.~D.,  {Oishi} J.~S.,  {Skory} S.,  {Skillman} S.~W.,
  {Abel} T.,   {Norman} M.~L.,  2011, \mn@doi [\apjs]
  {10.1088/0067-0049/192/1/9}, \href
  {http://adsabs.harvard.edu/abs/2011ApJS..192....9T} {192, 9}

\bibitem[\protect\citeauthoryear{{Vogelsberger}, {Sijacki}, {Kere{\v s}},
  {Springel}  \& {Hernquist}}{{Vogelsberger} et~al.}{2012}]{vogelsberger12}
{Vogelsberger} M.,  {Sijacki} D.,  {Kere{\v s}} D.,  {Springel} V.,
  {Hernquist} L.,  2012, \mn@doi [\mnras] {10.1111/j.1365-2966.2012.21590.x},
  \href {http://adsabs.harvard.edu/abs/2012MNRAS.425.3024V} {425, 3024}

\bibitem[\protect\citeauthoryear{{Vogelsberger} et~al.,}{{Vogelsberger}
  et~al.}{2014}]{vogelsberger14}
{Vogelsberger} M.,  et~al., 2014, \mn@doi [\mnras] {10.1093/mnras/stu1536},
  \href {http://adsabs.harvard.edu/abs/2014MNRAS.444.1518V} {444, 1518}

\bibitem[\protect\citeauthoryear{{Weinberger} et~al.,}{{Weinberger}
  et~al.}{2017}]{weinberger17}
{Weinberger} R.,  et~al., 2017, \mn@doi [\mnras] {10.1093/mnras/stw2944}, \href
  {http://adsabs.harvard.edu/abs/2017MNRAS.465.3291W} {465, 3291}

\makeatother
\end{thebibliography}

\bsp
\label{lastpage}
\end{document}